\documentclass[runningheads]{llncs}

\usepackage{cite}
\usepackage{amsmath,amssymb,amsfonts}
\usepackage{algorithmic}
\usepackage{graphicx}
\usepackage{textcomp}
\usepackage{xcolor}

\usepackage[T1]{fontenc}
\usepackage{rotating}
\usepackage{pbox}
\usepackage{url}
\usepackage{subfig}
\usepackage{breqn}
\usepackage{multirow}
\usepackage{verbatim}
\usepackage[normalem]{ulem}
\usepackage{algorithm}
\usepackage{caption}
\usepackage{subcaption}
\usepackage{booktabs}
\usepackage{array}

\begin{document}

\title{qGAMMA: A Framework for Optimal DNN Mapping on Multi-Precision Accelerators}

% \titlerunning{qGAMMA: A Framework for Optimal DNN Mapping on Multi-Precision Accelerators}

\titlerunning{\mbox{}}
\author{Leandro Fiorin\inst{1}\orcidID{0000-0003-2630-1509} \and
Luigi Altamura\inst{2}\thanks{This work was performed while at Politecnico di Milano.}\orcidID{0009-0000-2070-3932} \and
Cristina Silvano\inst{1}\orcidID{0000-0003-1668-0883}}
\authorrunning{L. Fiorin et al.}
\institute{Dipartimento di Elettronica e Informazione, Politecnico di Milano, Milano, Italy. \email{\{leandro.fiorin, cristina.silvano\}@polimi.it}\\
\and
Department of Computer Science and Engineering,
Chalmers University of Technology and University of Gothenburg, SE-41296 Gothenburg, Sweden\\
\email{altamura@chalmers.se}}

\maketitle

\begin{abstract}
%Deep Neural Network (DNN) accelerators integrate multiple processing elements that interact with the memory hierarchy via a network-on-chip. These accelerators' latency and energy efficiency depend heavily on optimizing the mapping of DNN algorithms to the available hardware resources, minimizing data movement, and enhancing data reuse. Adopting mixed-precision arithmetic in DNN computations has been shown to reduce latency, energy consumption, and memory requirements.
This work describes qGAMMA, a flexible framework for exploring and optimizing the mapping of mixed-precision Deep Neural Network (DNN) models onto general-purpose spatial accelerators that support a range of fixed- and floating-point precisions.
qGAMMA navigates the design space using a domain-specific genetic algorithm and an analytical performance model based on hardware synthesis results.
The framework identifies the optimal mapping and accelerator configuration that minimizes target metrics.
Extensive evaluations across DNN workloads show that qGAMMA offers valuable insights for designing mixed-precision applications, enabling better trade-offs between performance, energy efficiency, and resource utilization.

%The framework identifies the optimal mapping and accelerator configuration that minimizes the target optimization metric.
%Through extensive evaluation across several DNN inference workloads, we demonstrate that qGAMMA provides valuable insights for designing mixed-precision applications, helping designers achieve better trade-offs between performance, energy efficiency, and resource utilization.

% SHORTENED VERSION - NO space gain
%This work describes qGAMMA, a flexible framework for optimizing the mapping of DNN models onto spatial accelerators supporting various fixed- and floating-point precisions. 
%qGAMMA uses a domain-specific genetic algorithm and an analytical performance model to explore the design space, identifying optimal mappings and configurations that minimize target metrics.
%Extensive evaluations across DNN workloads show that qGAMMA offers valuable insights for designing mixed-precision applications, enabling better trade-offs between performance, energy efficiency, and resource utilization.

\keywords{DNN accelerators \and Optimization \and Mapping.}
\end{abstract}

%----------------------------------------------------
\section{Introduction}

%As DNNs grow more computationally demanding, efficient hardware accelerators have become crucial for DNNs, especially on resource-limited platforms.
%Recent research has explored low- and mixed-precision computation to reduce storage and computational demands on DNN accelerators.
%Each layer of a DNN can be processed with a different precision, offering additional flexibility in balancing efficiency and accuracy\cite{Sun-2020, Lee-2021, Yao-2021}.
%However, the optimization and fine-tuning of the execution of mixed-precision DNN models on accelerators with multiple precision pipelines (INT and/or FP) introduce complexities that remain largely unaddressed in system optimization.
%\footnote{In this work, we use the term \textit{mixed-precision} when discussing DNN models potentially using a different precision for the computations each of the layers,  while the term \textit{multiple-precision} for indicating the availability in the accelerator of multiple computing pipelines each computing on operands with a different precision.}.
As DNNs become increasingly computationally demanding, efficient hardware accelerators have become essential, particularly for resource-constrained platforms. 
Recent research has focused on low- and mixed-precision computation to reduce storage and computational overhead in DNN accelerators.
By processing each DNN layer with a different precision, these approaches offer greater flexibility in balancing efficiency and accuracy\cite{Sun-2020, Lee-2021, Yao-2021}. 
However, optimizing and fine-tuning the execution of mixed-precision DNN models on accelerators equipped with multiple precision pipelines (integer and/or floating point) introduces significant complexities largely unaddressed in system-level optimization.

%This work presents qGAMMA, a framework designed to optimize the mapping of mixed-precision DNN models on accelerators that support multiple precision pipelines.
%In particular, qGAMMA enables designers to explore the trade-offs between mapping, parallelization, and precision, providing insights on how to leverage mixed-precision computations for optimal performance.
%To the best of our knowledge, qGAMMA is the first tool specifically designed to address this optimization challenge.

This work introduces qGAMMA, a framework designed to optimize the mapping of mixed-precision DNN models onto accelerators that support multiple precision pipelines. Specifically, qGAMMA enables designers to explore trade-offs between mapping strategies, parallelization, and precision, offering valuable insights into leveraging mixed-precision computations for optimal performance. 
To the best of our knowledge, qGAMMA is the first tool specifically developed to tackle this optimization challenge.

In this work, we use the term \textit{mixed-precision} when discussing DNN models in which each layer is potentially computed using a different precision, while the term \textit{multi-precision} is used to indicate the availability in the accelerator of hardware modules supporting computations in different precisions.

The main contributions of this work are as follows:
\begin{itemize}
\item \textbf{qGAMMA}: A framework for optimizing the mapping of mixed-precision DNN models on multi-precision accelerators;
\item \textbf{Architectural template}: A configurable accelerator template that models state-of-the-art multi-precision architectures, included in qGAMMA to support the exploration process;
\item \textbf{Performance estimation tool}: An analytical tool for estimating the performance of multi-precision accelerators, based on the proposed architectural template;
\item \textbf{Heuristic search optimization}: The use of heuristics to prune the search space of qGAMMA, filtering out inefficient mappings and improving optimization efficiency;
\item \textbf{Experimental validation}: A comprehensive experimental campaign showing the effectiveness of qGAMMA in estimating and optimizing performance.
\end{itemize}

%To the best of our knowledge, qGAMMA is the first tool specifically designed to address the challenges of jointly optimizing DNN mapping and accelerator architecture for mixed-precision computations.
%Following the publication of this work, we plan to release qGAMMA as an open-source tool, enabling broader access to its capabilities and fostering further research in this area.
This paper is organized as follows. 
Section~\ref{sec:background} provides the relevant background and reviews the related work, while Section~\ref{sec:accelerator} describes the target accelerator architecture template.
Section~\ref{sec:qGAMMA} presents the key features of the qGAMMA framework.
Section~\ref{sec:experiments} discusses experimental evaluations, while Section~\ref{sec:conclusions} concludes the paper with final remarks.
\section{Background and related work}
\label{sec:background}

%-------------------------------------------
\begin{figure}[t!]
 \centering
 	\subfloat[][]{
	\includegraphics[width=0.48\columnwidth]{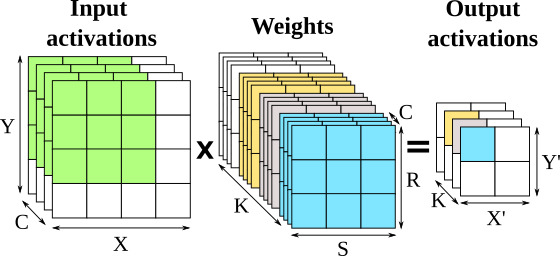}
	\label{fig:convolution}}
	\subfloat[][]{
	\qquad
    \includegraphics[width=0.28\columnwidth]{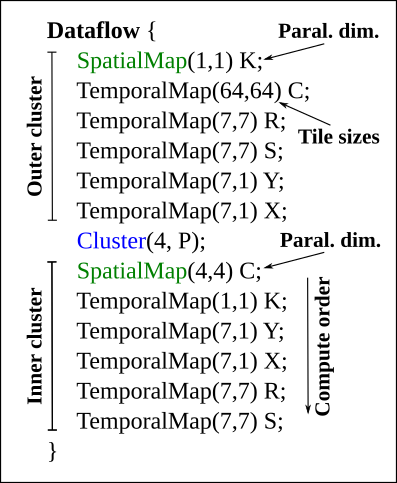}
	\label{fig:dataflow}}
    \caption{An example of (a) CNN computation and (b) dataflow structure.}
 \label{fig:conv_dataflow}	
\end{figure}
%-------------------------------------------

The 2D CNN\footnote{In this work, we will use the terms Deep Neural Networks (DNNs) and CNNs (Convolutional Neural Networks) interchangeably. We mainly focus on CNNs, as matrix-matrix and matrix-vector multiplications, common operations in DNNs, can be represented as convolutions by setting to one some of the CNN dimensions\cite{Parashar-2019}.} algorithm consists of six nested loops (or seven when considering batching) of Multiply-Accumulate (MAC) operations.
These loops can be ordered, tiled, and scheduled to optimize the algorithm execution on an accelerator.
This configuration, known as \textit{dataflow}, directly influences resource utilization, energy efficiency, and data reuse\cite{Kwon-2019}.
Dataflow is defined by (i) \textit{computation order}, affecting temporal and spatial reuse, (ii) \textit{parallelism and unrolling}, determining data movement patterns, and (iii) \textit{tiling strategy}, which dictates memory usage and bandwidth requirements\cite{Kao-2020}.

%-------------------------------------------
\begin{table}[t]
\renewcommand{\arraystretch}{1.3}
\centering
\caption{Comparison of mapping frameworks' characteristics in previous works and this work. qGAMMA is the first tool to support mixed-precision mapping on multi-precision spatial architectures.}\label{Tab:comp}
\begin{tabular}{|l|c|c|c|}
\hline
Framework & Exploration method & Mult. precision models & Mixed-precision\\
\hline
%Timeloop\cite{Parashar-2019} & Linear search/Random sampling & Yes & No \\
Timeloop\cite{Parashar-2019} & Random sampling & \textbf{Yes} & No \\

GAMMA\cite{Kao-2020} & Genetic algorithm & No & No \\
LOMA\cite{Symons-2021} & Constrained brute-force & No & No \\
CoSA\cite{Huang-2021} & Mixed ILP formulation & No & No \\
SALSA\cite{Jung-2023} & Simulated annealing & No & No \\
This work & Genetic algorithm + heuristics & \textbf{Yes} & \textbf{Yes} \\

\hline
\end{tabular}
\end{table}
%-------------------------------------------
State-of-the-art tools like MAESTRO\cite{Kwon-2019} and Timeloop\cite{Parashar-2019} adopt the dataflow concept to evaluate CNN performance on spatial accelerators. 
In MAESTRO, a 2D convolution and its dataflow model (Figure~\ref{fig:conv_dataflow}) are structured using key directives: \textit{TemporalMap()}, which dictates data distribution over time within a processing element (PE), \textit{SpatialMap()}, which indicates spatial distribution data across multiple PEs for parallel execution, and \textit{Cluster()}, which groups PEs or subclusters for multidimensional spatial distribution.
These tools analyze dataflow descriptions and accelerator configurations, using analytical models to estimate arithmetic unit activity, memory usage, communication overhead, and data reuse efficiency.
Execution latency is dictated by the slowest component, while energy consumption is estimated using activity counts and component-specific energy costs.

%GAMMA\cite{Kao-2020}, along with its extension DiGAMMA\cite{Kao-2022}, enables the exploration of the design space for per-layer dataflows tailored to specific hardware accelerators. 
%GAMMA uses a domain-specific genetic algorithm to explore optimal mappings of CNN layers to the target hardware, aiming to minimize key performance metrics such as latency and energy consumption.
%As shown in Figure~\ref{fig:l3-genome}, the tool encodes the dataflow characteristics of each cluster level as a set of gene pairs — one for each of the six dimensions of the CNN algorithm, along with an additional pair to specify the spatially parallelized dimension. 
%This genetic encoding directly translates into a dataflow model, such as the one shown in Figure~\ref{fig:dataflow}, which can be evaluated using a performance estimation tool.
%GAMMA generates new solutions by varying the size and position of dataflow directives, continuing the search until a predefined number of samples has been explored.

GAMMA\cite{Kao-2020} supports the exploration of per-layer dataflows optimized for specific hardware accelerators.
As shown in Figure~\ref{fig:l3-genome}, GAMMA encodes dataflow characteristics as gene pairs—one for each of the six CNN dimensions (see Figure~\ref{fig:convolution}) and an additional pair for spatial parallelization-and use a domain-specific genetic algorithm to search for an optimal CNN layer mappings minimizing latency and/or energy consumption.
By adjusting the size and position of dataflow directives, GAMMA iteratively refines solutions until a predefined search limit is reached.
The dataflows selected are used as input to performance estimation tools such as MAESTRO or Timeloop to evaluate the objective metrics.

Additional tools such as LOMA\cite{Symons-2021}, CoSA\cite{Huang-2021}, and SALSA\cite{Jung-2023} address the optimal mapping and scheduling of DNNs onto spatial accelerators, either through brute force exploration, mixed integer linear programming (ILP) formulations or by using simulated annealing to transform loops and loop orders.

Table~\ref{Tab:comp} compares the key characteristics of previous work and qGAMMA.
Existing frameworks support the exploration of the DNN mapping space but assume a single, generic, and unspecified precision for computations.
Among them, only Timeloop supports analytical estimation models that account for different possible precision values in single-precision computation pipelines.
qGAMMA is the first framework to explicitly incorporate mixed-precision into dataflow and mapping exploration, and it is specifically designed to address the challenge of optimizing the mapping of mixed-precision DNN models on multi-precision spatial architectures.
\section{Accelerator architecture template}
\label{sec:accelerator}

%-----------------------------------------------------------------
% picture with overview architecture
\begin{figure}[t]
 \centering
 \includegraphics[width=0.7\columnwidth]{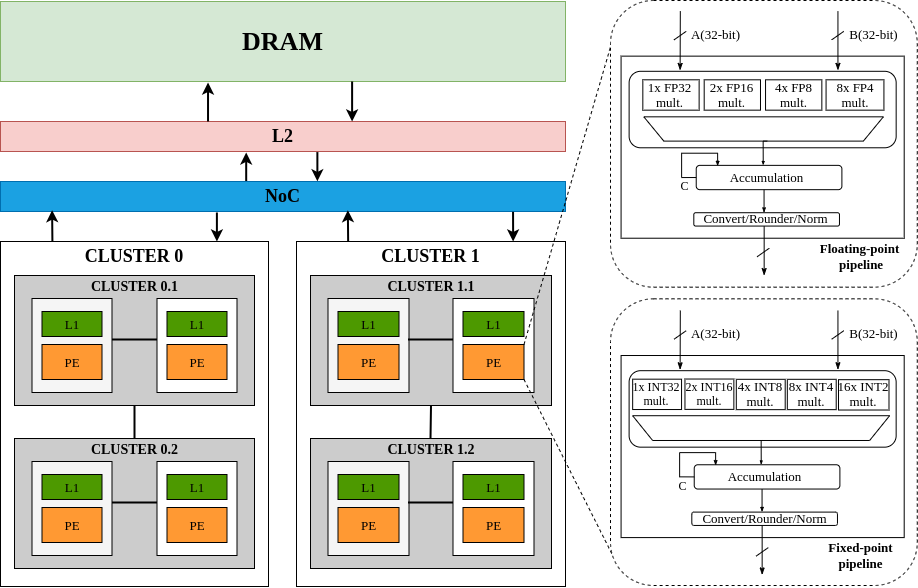}
 \caption{Target accelerator architecture template.}
  \label{fig:architecture}
\end{figure}
%-----------------------------------------------------------------

%Figure~\ref{fig:architecture} illustrates the spatial architecture template considered in this work.
%The architecture consists of a collection of PEs, each with a local scratchpad buffer (L1), a global scratchpad buffer (L2), a main memory, and a Network-on-Chip (NoC) connecting all components.
%We assume a double-buffering implementation for the accelerator's buffers to optimize data transfer efficiency.
%Depending on the design configuration, PEs can be organized into multiple levels of physical clusters.

Figure~\ref{fig:architecture} presents the spatial architecture template used in this work, which consists of PEs with local (L1) and global shared (L2) scratchpad buffers, main memory, and a Network-on-Chip (NoC) interconnecting all components. 
The PEs can be arranged into multiple hierarchical clusters.
To improve data transfer efficiency, the accelerator utilizes double buffering. 

To support mixed-precision computations, each PE in our model features a configurable functional unit based on state-of-the-art designs\cite{Andersch-2022, Lee-2021} and capable of performing computations in different precisions.
Following\cite{Lee-2021}, we consider separate pipelines for integer fixed-point and floating-point operations, allowing independent optimization for improved energy efficiency, although with added area overhead.
Functional units have 32-bit input/output ports and support FP32, FP16, FP8, and FP4 floating-point operations\cite{Sami-2024, Sun-2020}, as well as INT32, INT16, INT8, INT4, and INT2 fixed-point computations\cite{Lee-2021, Jouppi-2021}. 
At FP32 or INT32 precision, the functional unit performs a single MAC operation.
For lower precisions $Q$, the 32-bit lane is partitioned into $P = 32/Q$ operands and the functional unit executes a $\sum_{P}^{} A_{i} \cdot B_{i} + C$ operation\cite{Lee-2021, Andersch-2022}. 
As in \cite{Lee-2021}, integer fixed-point and floating-point multipliers use a different number of pipeline stages.
Given the high parallelism of the targeted algorithms, threads are scheduled to minimize data dependencies, ideally achieving one instruction per cycle\cite{Jouppi-2021}.
During design-space exploration, accelerator components are instantiated automatically with precision-specific configurations, including accumulator width, adder trees, and output packing.
\section{qGAMMA}
\label{sec:qGAMMA}

%-----------------------------------------------------------------
\begin{figure}[t]
 \centering
 \includegraphics[width=0.6\columnwidth]{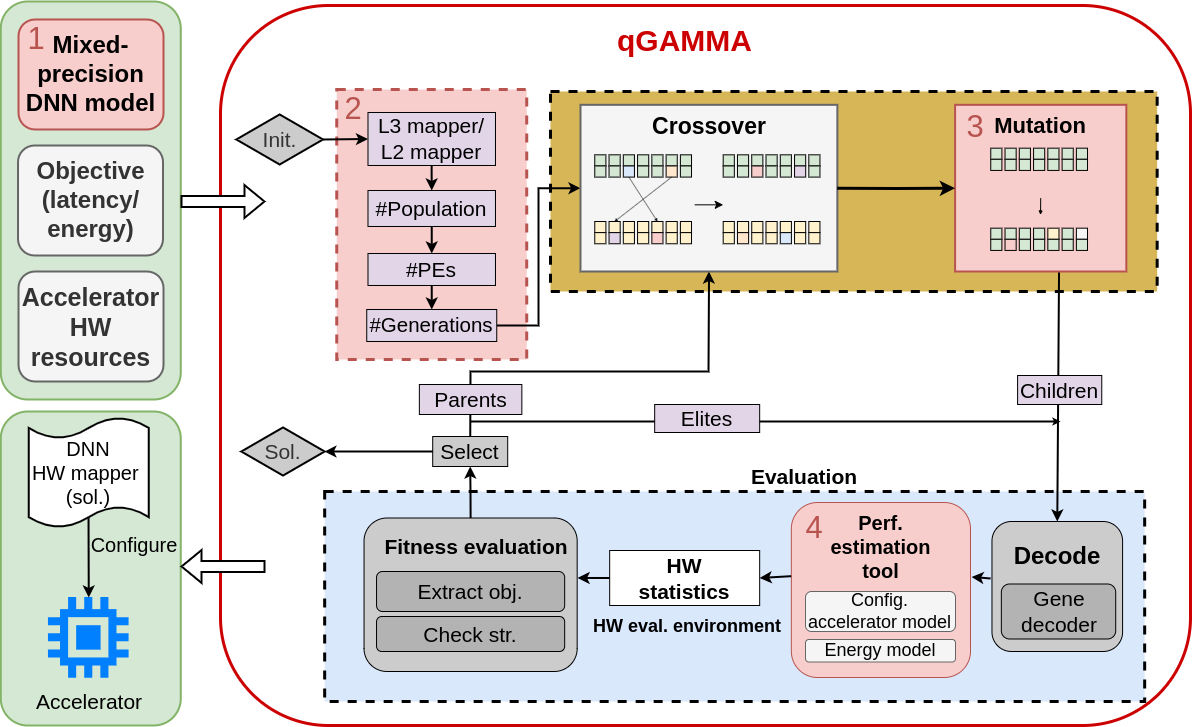}
 \caption{Overview of the proposed framework: The red boxes highlight the key technical contributions of this study.}
 \label{fig:qGAMMA}
\end{figure}
%-----------------------------------------------------------------
As shown in Figure~\ref{fig:qGAMMA}, qGAMMA takes as input a mixed-precision DNN model, an objective metric (energy, latency, or energy-delay product (EDP)), and a general description of accelerator resources (such as the number of PEs, memory levels, maximum memory sizes, maximum NoC bandwidth, and supported precisions).
By leveraging the search based on a genetic algorithm and the dataflow encoding used in GAMMA\cite{Kao-2020}, qGAMMA explores a range of possible dataflows and architectural configurations—such as the number of components in each hardware cluster and buffer utilization across memory levels— to identify the characteristics that minimize the chosen metric.

Figure~\ref{fig:qGAMMA} shows the components of qGAMMA. 
Our key contributions, highlighted in red in the figure, are as follows:
\begin{enumerate}
    \item \textbf{Enhanced dataflow model}: We extended the dataflow model to account explicitly for the precision of operands in each DNN layer;
    \item \textbf{Enhanced genetic encoding}: We extended the genetic encoding to support an additional cluster level, enabling the modeling of the multi-precision pipelines of the proposed architectural template;
    \item \textbf{Heuristic optimization}: To mitigate the increased design space resulting from the larger genome due to the additional cluster level needed to model the behavior of the multi-precision pipeline, we implemented several heuristics to exclude inefficient dataflows and reduce exploration time; 
    \item \textbf{Performance estimation tool}: Leveraging MAESTRO\cite{Kwon-2019}, we developed an analytical tool to assess performance and energy consumption, accounting for mixed-precision computations on the target architecture.
\end{enumerate}

%-------------------------------------------
\begin{figure}[t!]
 \centering
 	\subfloat[][]{
	\includegraphics[width=0.6\columnwidth]{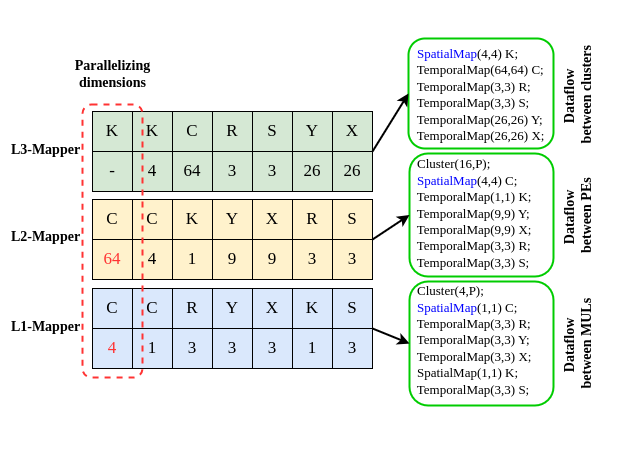}
	\label{fig:l3-genome}}
    \qquad
	\subfloat[][]{
	\includegraphics[width=0.3\columnwidth]{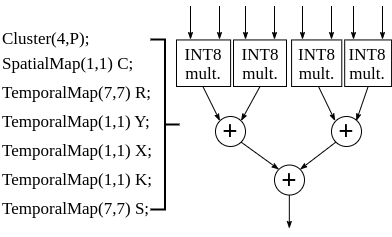}
	\label{fig:dataflow_pipe}}
    \caption{(a) Example of encoding for a 3-level genome and corresponding dataflow; (b) dataflow modelling of an INT8 operation in the proposed functional unit model.}
 \label{fig:genoma}	
\end{figure}
%-------------------------------------------

\subsection{Enhanced dataflow model}
We enhanced the GAMMA framework to support mixed-precision DNN models. In addition to specifying the dimensions of each DNN layer, hardware constraints, and the objective metric, qGAMMA allows users to define the precision of weights and activations for each layer. 
This information is then used to generate the architectural model of the pipeline and the corresponding dataflow model used in the exploration.

\subsection{Extended genetic encoding}
\label{ssec:ext_gen_enc}
qGAMMA encodes the dataflow of each DNN layer as a genome for the genetic algorithm driving the exploration, with a separate section for each cluster level. 
To integrate our multi-precision functional unit into qGAMMA, we model the $\sum_{P}^{} A_{i} \cdot B_{i} + C$ operation in terms of dataflow directives (Figure~\ref{fig:dataflow_pipe}).
A low-precision pipeline is modeled as an additional inner cluster with a size equal to $P$, where spatial mapping along the CNN's output channel dimension ($K$ in Figure~\ref{fig:convolution}) is restricted.
This is to prevent the mixing of partial sums from different output channel calculations in the intermediate sums of the functional unit.
As shown in Figure~\ref{fig:dataflow_pipe} for INT8 precision, four multipliers operate in parallel over data read from the L1 scratchpad of the PE to compute the $\sum_{P}^{} A_{i} \cdot B_{i} + C$ instruction.
In this case, the functional unit is modeled as a cluster of four elements, and the related dataflow restricts spatial parallelization over the $K$ dimension.

To support the functional unit's architectural model, we extended GAMMA’s original 2-level genome (one genome level for each cluster level shown in Figure\ref{fig:architecture}) by introducing an additional level of parallelism, accounting for the cluster used to model the low-precision pipeline. 
Figure~\ref{fig:l3-genome} illustrates the encoding of the 3-level genome:
The first block (L3-mapper) defines the dataflow between PE clusters, the second block (L2-mapper) represents the dataflow within each PE cluster, and the third block (L1-mapper) models the dataflow for the multi-precision pipeline.
As in GAMMA, within each block, the last six pairs of genes encode a DNN layer dimension and its corresponding tile size, ordered according to the computation sequence in the dataflow. 
The first pair of genes in each mapper specifies the parallelized dimension.
Their associated value is constrained by the corresponding tile size of the chosen parallelizing dimension and by the number of available elements on which the parallelization is performed (clusters, PEs, multipliers).

The 3-level genome is used exclusively for configurations involving low-precision computations, while FP32 and INT32 computations continue to use the original 2-level genome.
The inner cluster size is dictated by the lowest precision used in the DNN computations. 
In contrast, the sizes of higher-level clusters are dynamically selected during exploration based on the model’s overall performance.

qGAMMA’s genetic algorithm follows the same fundamental steps as the original framework\cite{Kao-2020}. 
It begins by generating a random initial population of samples (dataflows), then evolves the genome through crossover between two high-performing parent genomes. Additionally, with a certain probability, it applies random mutations to the parallel dimensions and tile sizes specified in the genome.
Once a population is selected, the genomes are decoded into dataflow descriptions and evaluated using a hardware performance estimation tool.
This tool assesses the objective function while ensuring compliance with hardware constraints. 
The next generation of samples is then selected based on their fitness, as determined by the chosen objective metric.

\subsection{Heuristic optimization}
Introducing a three-level genome significantly expands the search space compared to a two-level genome (size of $\mathcal{O}(10^{33})$ for a ResNet18\cite{Kao-2020}).
To prune the search space, we define a combination of three heuristics:
\begin{itemize}
    \item Heuristic 1: Since accelerator architectures typically consist of a power-of-two number of processing units, we restrict the number of PEs per cluster to powers of two. 
    An exception is made for the inner cluster, where the number of low-precision multipliers determines the number of PEs;

    \item Heuristic 2: The 3-level genome enables parallelization across three dimensions, one for each cluster level, prioritizing dimensions with more elements, as they are more likely to benefit from parallelization. The outer cluster randomly selects the parallel dimension from the three largest dataflow dimensions. At the next cluster level, the selection is narrowed to the two largest dimensions. No restrictions are imposed on the inner cluster, as the parallelization over the K dimension is already constrained;

    \item Heuristic 3: We explore power-of-two values for the K and C dimensions during genome generation since these values typically correspond to the most common configurations in DNN layers.
\end{itemize}

\subsection{Performance estimation tool}
Leveraging MAESTRO\cite{Kwon-2019}, we designed an analytical estimation tool supporting multi-precision pipelines and incorporating the changes introduced in qGAMMA.
We modified the original dataflow description used as input to the estimation tool to include the desired computation precision for each DNN layer.
To reflect the architectural features of the PEs introduced in Section~\ref{sec:accelerator}, we developed a new activity-based energy model that accounts for the implementation of low-precision MACs, packed accesses to memory buffers, and NoC transfers.
More details on the energy model are provided in Section~\ref{sec:experiments}.
\section{Experimental evaluation}
\label{sec:experiments}

\subsection{Experimental setup}

\subsubsection{Hardware accelerator model}
Our experiments focus on a hardware accelerator consisting of 16 PEs, each with a maximum of 512 bytes of L1 scratchpad memory, and a global L2 scratchpad memory of up to 512 Kbytes.
Energy, performance, and area values for the functional unit components were extracted from implementation results published in \cite{Lee-2021, Jouppi-2021, Sun-2020}.
The on-chip interconnection network is modeled based on implementation results from \cite{Chen-2014}. 
Energy and area values for the memory elements were obtained using CACTI-D \cite{Thoziyoor-2008}.
All values were scaled to 22-nm technology.
In our results, we did not include energy consumption for main memory, as the exploration framework discards design solutions requiring main memory access for storing temporary results, resulting in the same main memory energy profile across all valid solutions.

\subsubsection{DNN models}
%\label{sssec:dnn_models}
We evaluated three DNN models of different complexity: a MobileNetV1\cite{Howard-2017}, a ResNetV1\cite{He-2016}, and a mixed-precision variant of ResNet18\cite{Yao-2021}, which employs a combination of INT8 and INT4 precisions for weights and activations.
In\cite{Yao-2021}, the selection of layers to be quantized in INT4 is formulated as an ILP problem aiming at minimizing the DNN's sensitivity to perturbations caused by the lower precision while adhering to constraints on latency, computing capacity, and model size.
The resulting mixed-precision DNN model is then fine-tuned, achieving an accuracy close to the FP32 baseline (-1.5\%) and a uniform INT8 ResNet18 (-1.6\%). 
The precision assigned to each layer is shown in Figure~\ref{fig:heuristics_latency}.

\subsubsection{Optimization setup}
%qGAMMA can optimize up to two metrics per execution. 
%We focused on single-objective optimization, targeting either latency, activity-based energy, or EDP.
%
%We used qGAMMA in two distinct optimization approaches.
%In the first approach, we performed per-layer optimization, enabling a distinct clustering strategy for each DNN layer. 
%To achieve this, we ran qGAMMA for 1000 generations with a population of 150 individuals, using INT32 and FP32 precisions, which require a 2-level mapper. 
%For lower precision configurations requiring a 3-level mapper, we increased the population size to 300 individuals, as the search space grows significantly for these configurations.
%
%In the second approach, we considered a single clustering that applies to all layers, reflecting that the clustering is a hardware configuration that cannot be altered at runtime. 
%In this case, we first ran the per-layer optimization as described in the first approach, then conducted a second phase in which we ran qGAMMA for 1000 generations, selecting a single clustering based on the configuration that was most frequently used across all layers during the first phase.
%
%To improve the efficiency of the optimization process, we implemented an early-stopping policy that halts the optimization if the best solution does not change after a specified number of generations (150 generations in our experiments).
%
%%%%
We focused on single-objective optimization, targeting either latency, activity-based energy, or EDP. 
We applied qGAMMA in two optimization approaches.
In the first approach, we performed per-layer optimization, allowing a unique clustering strategy for each DNN layer. 
We ran qGAMMA for 1000 generations with a population of 150 individuals, using INT32 and FP32 precisions (requiring a 2-level mapper).
For lower-precision configurations (requiring a 3-level mapper), we increased the population to 300 individuals due to the larger search space.

In the second approach, we used a single clustering applied to all layers, reflecting the fact that clustering is a hardware configuration fixed at runtime.
We first ran the per-layer optimization as in the first approach, then conducted a second phase where qGAMMA was run for 1000 generations to select a single clustering based on the configuration most frequently used across layers in the first phase.
To improve optimization efficiency, we implemented an early-stopping policy that halts the process if the best solution does not change after 150 generations.

%----------------------------------------------------------------
%----------------------------------------------------------------
\subsection{Experimental Results}
A direct comparison of qGAMMA with state-of-the-art mapping and exploration tools is not feasible, as none of the existing frameworks support both mixed-precision DNN modeling and multi-precision accelerators. 
However, this section presents experimental results that highlight qGAMMA’s capabilities, in comparison, when suitable, to the original search algorithm and modeling features of GAMMA\cite{Kao-2020}.
%------------------------------------------------------------------------
\subsubsection{Heuristics validation - ResNet18\cite{Yao-2021}}
%------------------------------------------------------------------------
%-------------------------------------------
\begin{figure}[t!]
 \centering
 	\subfloat[]{
	\includegraphics[width=0.75\columnwidth]{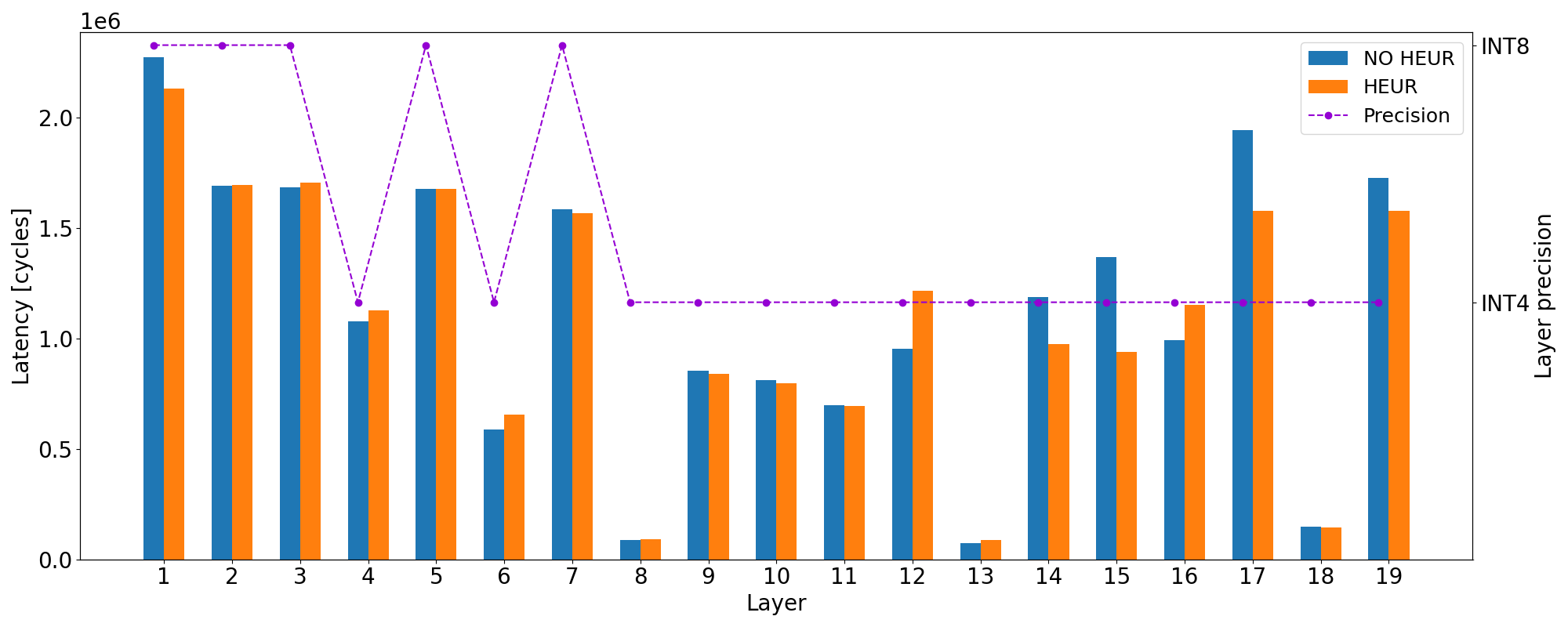}
	\label{fig:heuristics_latency}} \\
	\subfloat[]{
	\includegraphics[width=0.75\columnwidth]{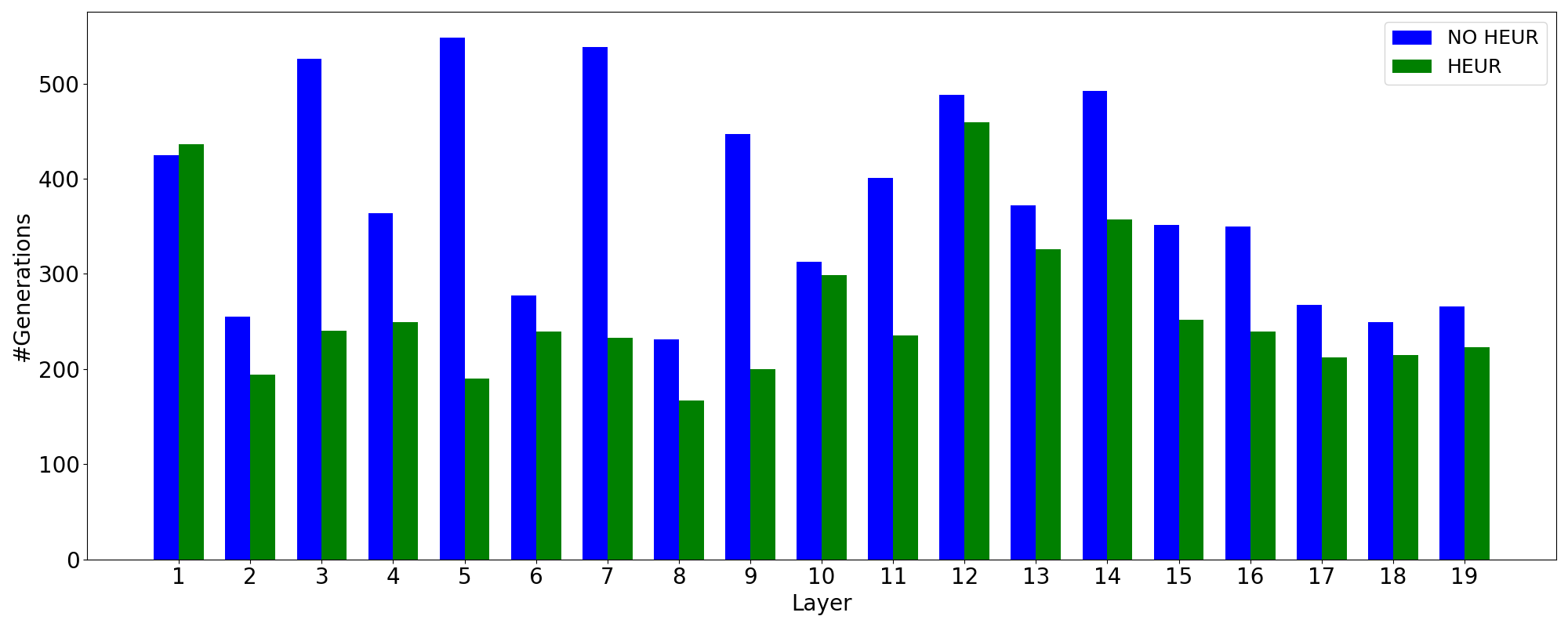}
	\label{fig:heuristics_generations}}
    \caption{(a) Latency and (b) number of generations observed for the mixed-precision ResNet18 optimized for minimum latency, when using the heuristics implemented in qGAMMA, and without them (original GAMMA search algorithm). The graphs include the generations needed to implement the early-stop. Per-layer precision is shown in Figure~\ref{fig:heuristics_latency}.}
 \label{fig:im2col-exe}	
\end{figure}
%-------------------------------------------

The proposed heuristics aim to prune the search space and eliminate inefficient dataflow. 
To demonstrate their effectiveness, we optimized the mixed-precision ResNet18 for minimum latency.

We ran qGAMMA with a population of 300 individuals for 1000 generations, applying an early-stop condition at 150 generations.
Figure~\ref{fig:heuristics_latency} compares the latency of the ResNet18 when using the heuristics implemented in qGAMMA, and without them (original GAMMA search algorithm).
The latency results are nearly identical in both cases, with a modest 4\% reduction when heuristics are applied. 
However, the heuristics significantly reduced the number of generations required to reach the early-stop condition and decreased overall execution time by about 31\% when considering the 150 early-stop generations (Figure~\ref{fig:heuristics_generations}), and by 51\% without them (results not shown). 
While slightly increasing the search time, the early-stop condition is essential for ensuring the convergence of the search algorithm.
%------------------------------------------------------------------------
\subsubsection{Uniform precision - ResNetV1\cite{He-2016}}

%-------------------------------------------
\begin{figure}[t!]
 \centering
 	\subfloat[]{
	\includegraphics[width=0.75\columnwidth]{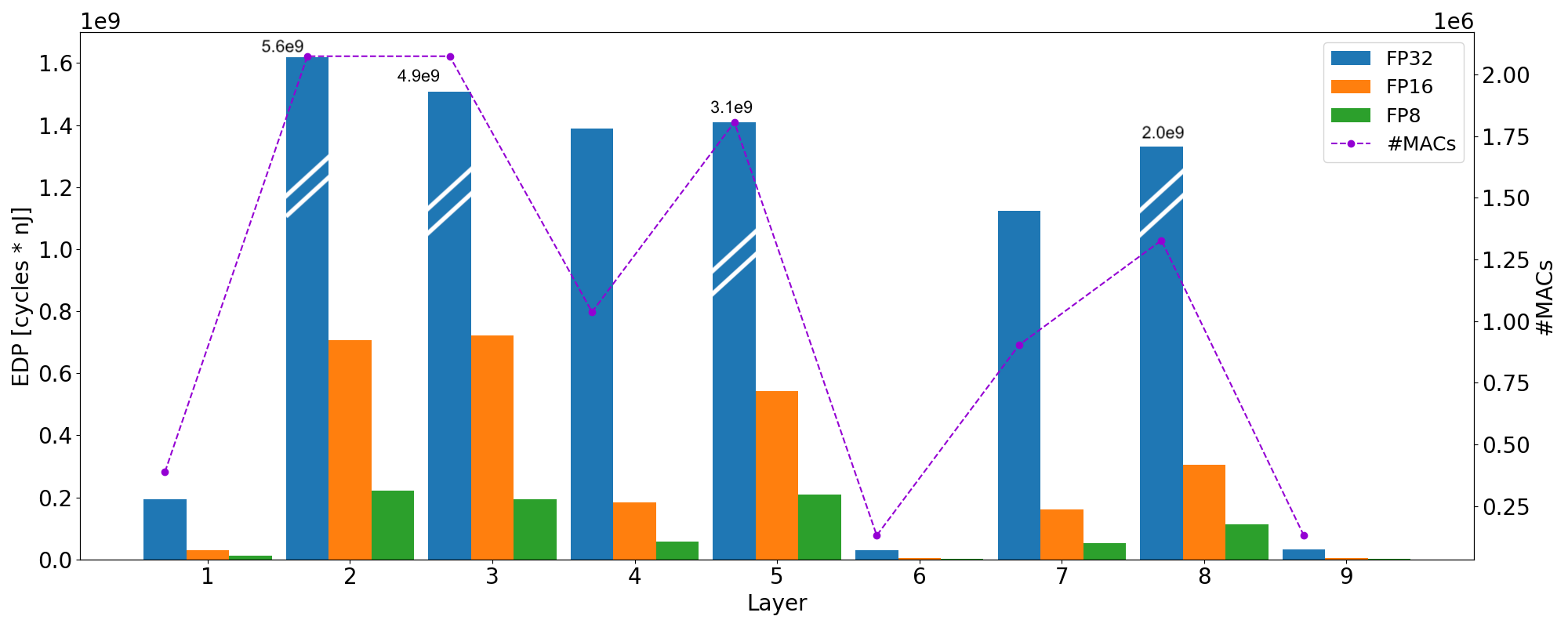}
	\label{fig:edp_resnetv1}} \\
	\subfloat[]{
	\includegraphics[width=0.75\columnwidth]{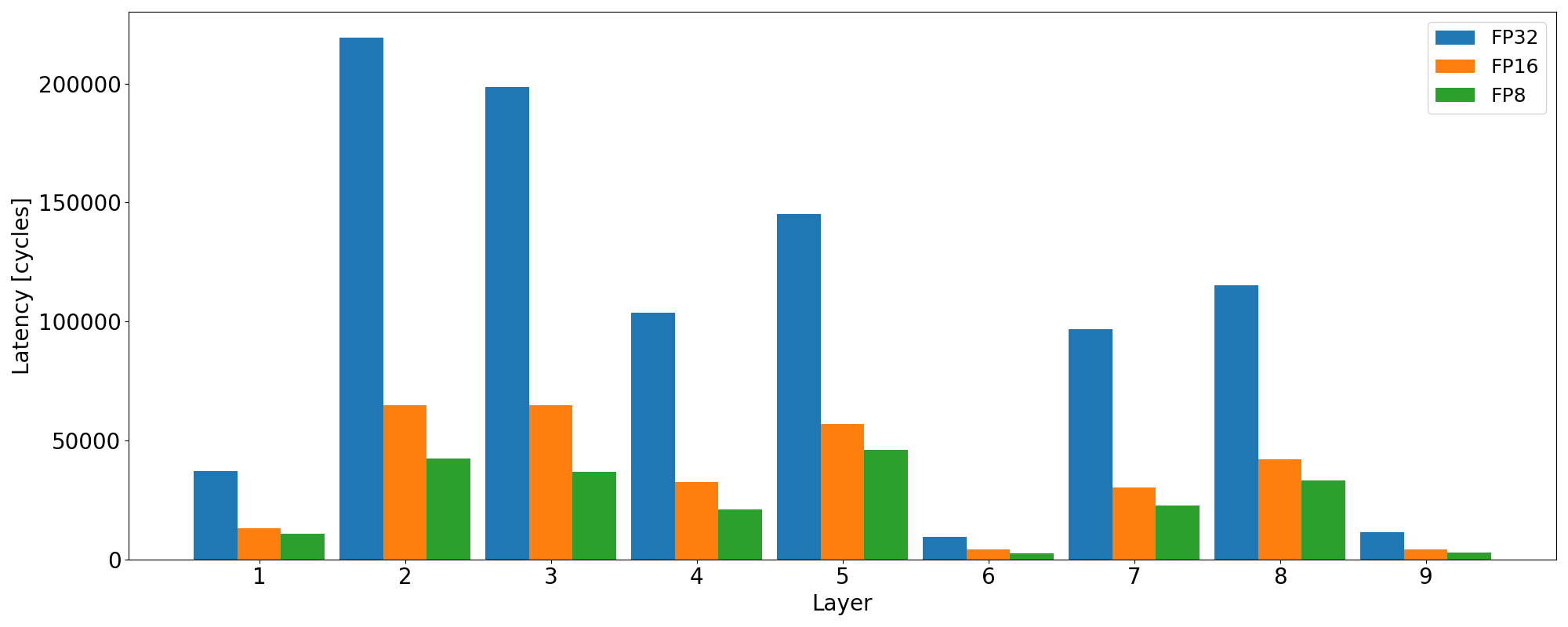}
	\label{fig:latency_resnetv1}} \\
%	\subfloat[]{
%	\includegraphics[width=\columnwidth]{images/activity_energy_resnetv1.png}
%	\label{fig:energy_resnetv1}} 
    \caption{(a) EDP and (b) latency of a ResNetV1 optimized for minimum EDP, considering three uniform floating-point precisions and per-layer optimization.}
 \label{fig:opt_resnetv1}	
\end{figure}
%-------------------------------------------

%In Figure~\ref{fig:opt_resnetv1}, we compare the results obtained when optimizing the EDP of ResNetV1 using three different uniform precision settings: FP8, FP16, and FP32. 
%For this experiment, we applied per-layer optimization.
%
In Figure~\ref{fig:opt_resnetv1}, we compare the results of optimizing with qGAMMA the EDP of ResNetV1 under three uniform precision settings: FP8, FP16, and FP32. 
This experiment follows a per-layer optimization approach.
The number of MAC operations per layer is displayed alongside the EDP.
As expected, lower precision leads to a reduced EDP, mainly due to lower latency from increased parallelization — enabled by the greater number of multipliers available in low-precision operations — and decreased energy consumption due to smaller operand bit-widths in computation, storage, and data transfer.
Switching from FP32 to FP16 improves EDP by 6.96$\times$, while transitioning from FP32 to FP8 results in a 21.44$\times$ improvement. 
A straightforward estimation might suggest a 4$\times$ gain — 2$\times$ from doubling available MAC units and 2$\times$ from halving energy costs in computation, buffering, and communication. 
However, the observed improvements surpass this estimate, driven by additional factors such as increased algorithmic parallelization, more energy-efficient operand accumulation, and reduced buffer and communication accesses due to fewer temporary accumulation reads/writes.

Figure~\ref{fig:latency_resnetv1} shows per-layer latency reductions: FP16 achieves approximately 3.0$\times$ lower latency compared to FP32, while FP8 reduces latency by 4.29$\times$.
Energy consumption (not shown) decreases by 2.32$\times$ when moving from FP32 to FP16 and by 4.94$\times$ when transitioning from FP32 to FP8.

%------------------------------------------------------------------------
\subsubsection{Uniform precision - MobileNetV1\cite{Howard-2017}}
%------------------------------------------------------------------------

%-------------------------------------------
\begin{figure}[t!]
 \centering
 	\subfloat[]{
	\includegraphics[width=0.75\columnwidth]{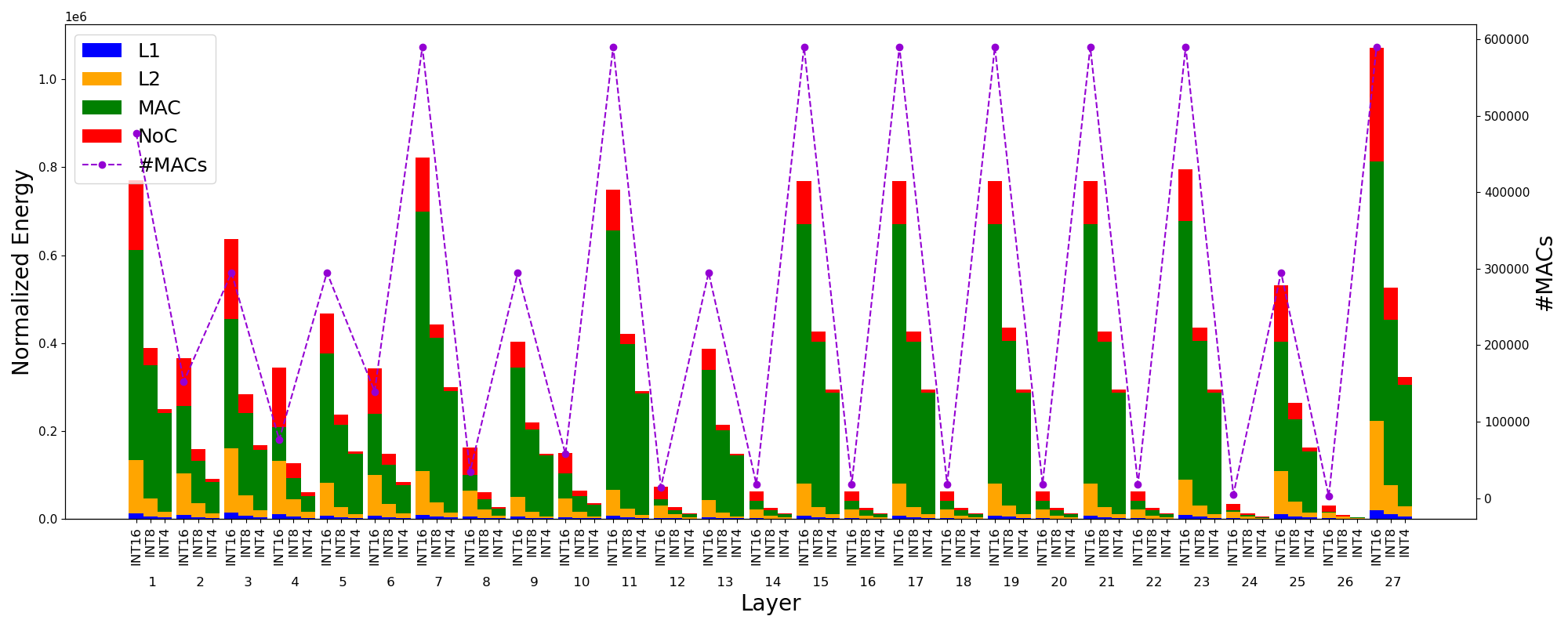}
	\label{fig:energy_mobilenet}} \\
	\subfloat[]{
	\includegraphics[width=0.75\columnwidth]{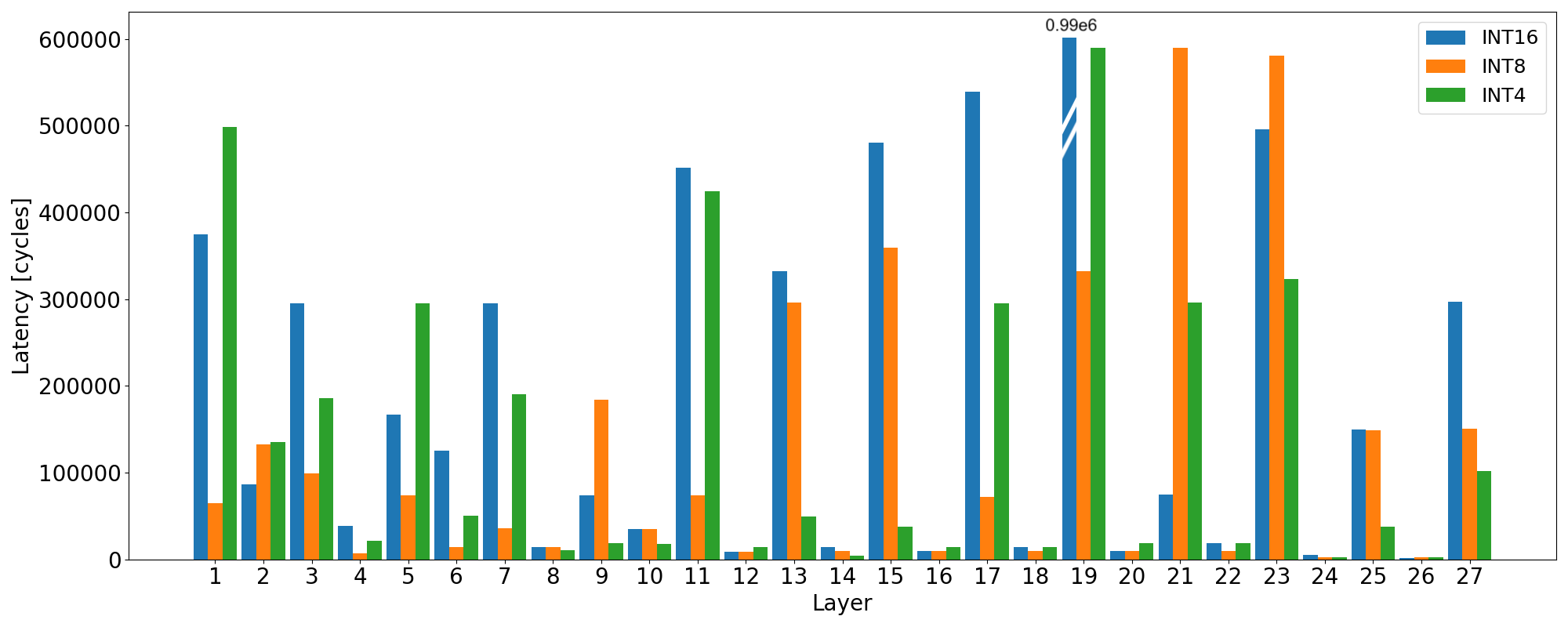}
	\label{fig:latency_mobilenet}} \\
    \caption{(a) Energy and (b) latency of a MobileNetV1 optimized for minimum energy, considering three uniform integer fixed-point precisions and single-cluster optimization.}
 \label{fig:opt_mobilenetv1}	
\end{figure}
%-------------------------------------------

%Figure~\ref{fig:opt_mobilenetv1} shows the results obtained when optimizing MobileNetV1 for energy consumption.
%In this experiment, we applied optimization with a single clustering for all layers, reflecting the experimental setup used when designing a new accelerator.
%We explored three uniform fixed-point precisions: INT16, INT8, and INT4.
%
Figure~\ref{fig:opt_mobilenetv1} presents the results of optimizing MobileNetV1 for energy efficiency using qGAMMA by setting the same clustering for all layers, mirroring the setup used in accelerators' design.
We evaluate three uniform fixed-point precisions: INT16, INT8, and INT4.

Figure~\ref{fig:energy_mobilenet} shows the per-layer energy breakdown, normalized to INT16 MAC energy, alongside the executed MAC operations. 
The primary energy cost arises from the computation pipelines, while memory buffers contribute minimally due to the small buffer sizes selected by qGAMMA and the dataflow's efficient data reuse strategies that reduce memory accesses. 
As expected, energy consumption scales with the precision, yielding reductions of 1.96$\times$ (INT16 to INT8) and 3.03$\times$ (INT16 to INT4).
The deviation from theoretical scaling is attributed to the fixed-point pipeline’s non-ideal energy scalability, and particularly to the reduced influence of the accumulation tree on the PE energy.

Figure~\ref{fig:latency_mobilenet} shows per-layer latency, revealing that lowering the precision does not always imply a reduced latency.
Since the optimization prioritizes energy minimization, the genetic algorithm may select configurations with suboptimal parallelization across PEs, leading to increased latency despite achieving minimal energy consumption.

%------------------------------------------------------------------------
\subsubsection{Comparison between mixed and uniform precision - ResNet18\cite{Yao-2021}}
%------------------------------------------------------------------------

%-------------------------------------------
\begin{figure}[t!]
 \centering
 	\subfloat[]{
	\includegraphics[width=0.75\columnwidth]{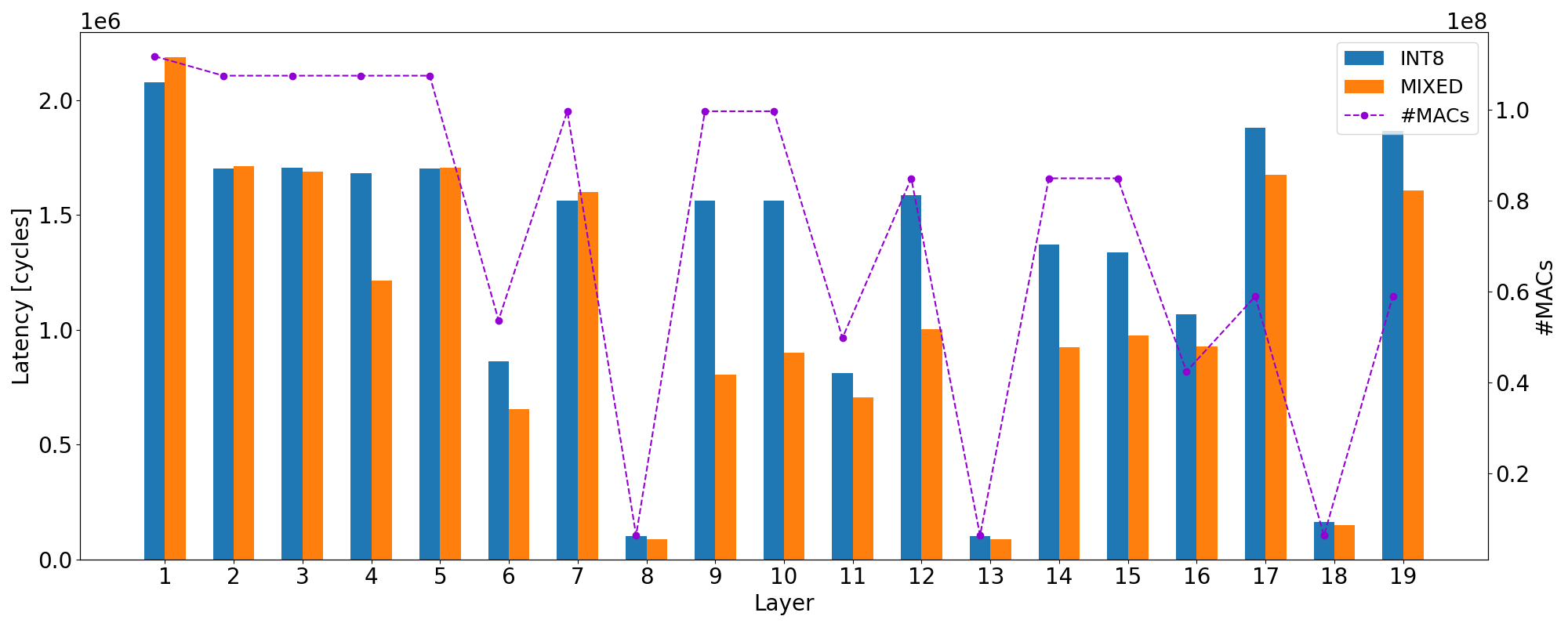}
	\label{fig:latency_resnet18}} \\
	\subfloat[]{
	\includegraphics[width=0.75\columnwidth]{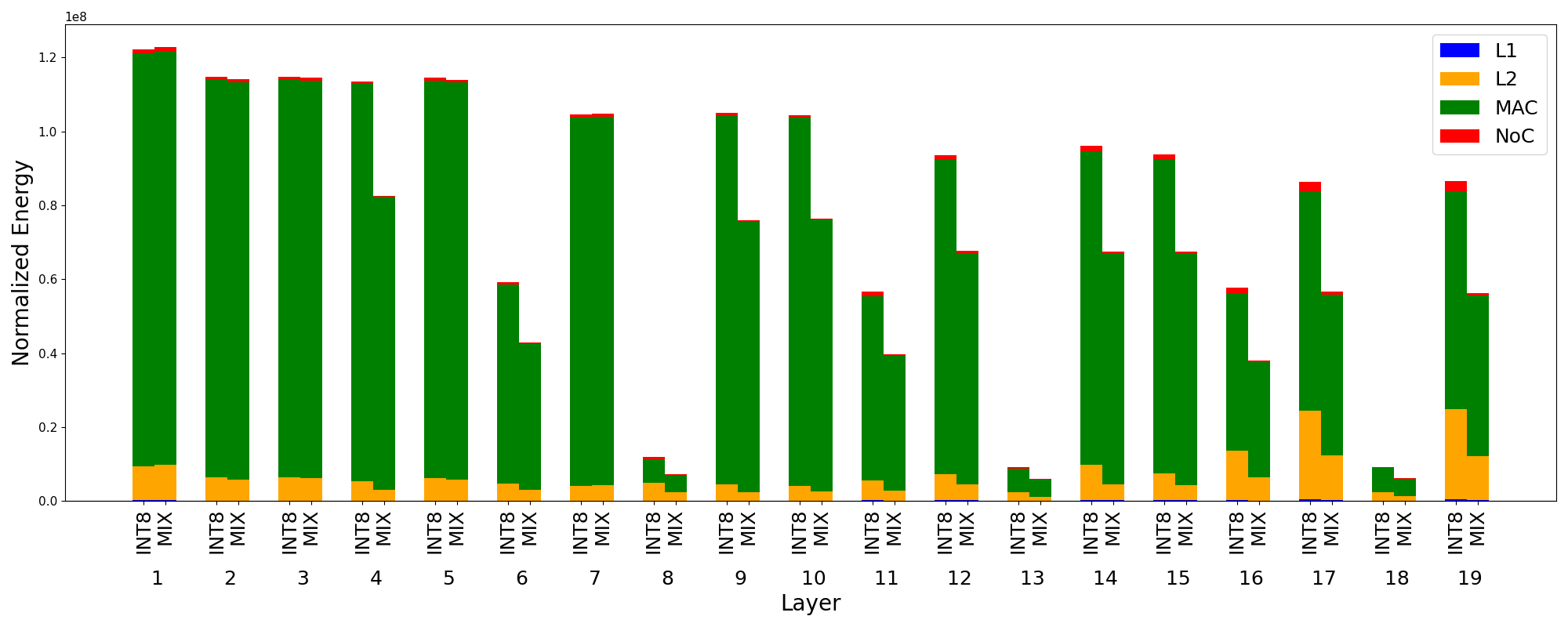}
	\label{fig:energy_resnet18}} \\
    \caption{Comparison of (a) latency and (b) energy consumption between a mixed-precision (INT8, INT4) ResNet18 and a single-precision (INT8) ResNet18, both optimized for minimum latency with single clustering optimization.}
 \label{fig:opt_resnetv18}	
\end{figure}
%-------------------------------------------

To evaluate the impact of mixed-precision models on mapping compared to a baseline with uniform quantization, we compared a mixed-precision (INT8, INT4) ResNet18 with a uniform INT8 ResNet18. 
In both cases, we optimized for minimum latency and applied a single clustering scheme to all layers.

Figure~\ref{fig:latency_resnet18} presents the per-layer latency and MAC operations. 
Key observations include:
\begin{itemize}    
%\item In this experiment, qGAMMA selects the clustering found in the layers with INT4 precision as optimal for minimizing the DNN latency. 
%However, this choice may result in a suboptimal configuration for INT8 layers (e.g., layers 1, 2, 3, 5, and 7).
%
\item qGAMMA selects the clustering found in INT4 layers as optimal for minimizing latency. 
However, this may lead to suboptimal configurations for INT8 layers (e.g., layers 1, 2, 3, 5, and 7).
\item Latency reductions depend on the specific characteristics of each layer and the extent to which parallelization can leverage available computational resources.
\item The genetic algorithm may occasionally converge to suboptimal solutions.
\end{itemize}
Overall, we achieve a 1.20$\times$ latency speedup, which aligns with the average improvement reported in \cite{Yao-2021} when comparing the mixed-precision and uniform INT8 ResNet18. 
Additionally, we observe a 17\% reduction in energy consumption.
\section{Conclusions}
\label{sec:conclusions}
%
%In this work, we proposed qGAMMA, a flexible framework designed for the optimization of DNN model mappings on accelerators supporting mixed precision, along with an accelerator architecture template that models state-of-the-art mixed-precision architectures.
%Through experimental evaluation across various DNN inference workloads, we demonstrated that qGAMMA can provide designers with valuable insights for optimizing mixed-precision applications.
%The framework offers estimation results that guide the selection of optimal configurations.
%With the inclusion of efficient techniques for quickly estimating DNN accuracy across varying models and precision levels, qGAMMA emerges as a powerful tool for the comprehensive optimization of DNN accelerators, enabling faster and more effective design-space exploration.
%
In this work, we introduced qGAMMA, a versatile framework for optimizing the mapping of mixed-precision DNN models on accelerators with multi-precision support. 
Alongside qGAMMA, we proposed an accelerator architecture template that models cutting-edge multi-precision designs.
Through extensive evaluation across diverse DNN inference workloads, we demonstrated that qGAMMA equips designers with valuable insights for optimizing mixed-precision applications.
By integrating efficient techniques for rapid DNN accuracy estimation across different models and precision levels, qGAMMA serves as a powerful tool for comprehensive DNN accelerator optimization, enabling faster and more effective design-space exploration.

%With the support of techniques for rapidly estimating the impact of varying DNN model precision on accuracy\cite{Wang-2020, Lin-2021}, qGAMMA offers a powerful tool for the design and optimization of edge-oriented DNN accelerators.

% ---- Bibliography ----
\bibliographystyle{splncs04}
\bibliography{biblio_short}

\end{document}